\documentclass[sigconf]{acmart}

\usepackage{booktabs} % For formal tables

\usepackage{algpseudocode}
\usepackage{mathtools}
\usepackage{algorithm}
\usepackage{xcolor}
\usepackage[subtle]{savetrees}
\usepackage{colortbl}
\usepackage{graphicx}
\usepackage{subcaption}
\usepackage{tikz-qtree}
\usetikzlibrary{decorations.text}
\usepackage{multirow}
\usepackage{microtype}
\usepackage{multicol}
\usepackage{pgfplots}
\usepackage{tabularx} % Tables spanning the \linewidth or 
\usepackage{graphics}
\usepackage{adjustbox}
\usepackage{csquotes}
\usetikzlibrary{pgfplots.groupplots}
\usepackage{mathrsfs,amsmath} 
\usepackage{nicefrac}
\pgfplotsset{compat=1.11}
\usepgfplotslibrary{ternary}
\usepackage{tikz}
\usetikzlibrary{calc,spy,shapes}
\usetikzlibrary{arrows,decorations.pathmorphing,fit,positioning,patterns}
\pgfplotsset{compat=newest}

\usepackage{setspace}
\newcommand{\shrink}{\vspace{-1.5ex}}
\newcommand{\sshrink}{\vspace{-.80ex}}

\def\:{\hskip0pt} % \: for hyphenation xxx\:---\:xxx 
\newcommand{\mypar}[1]{\vspace*{-0.2ex}\medskip\noindent\textbf{#1}~}
\newcounter{todocnt}

\DeclareFontFamily{U}{mathb}{\hyphenchar\font45}
\DeclareFontShape{U}{mathb}{m}{n}{
<-6> mathb5 <6-7> mathb6 <7-8> mathb7
<8-9> mathb8 <9-10> mathb9
<10-12> mathb10 <12-> mathb12
}{}
\DeclareSymbolFont{mathb}{U}{mathb}{m}{n}

\DeclareMathSymbol{\smalltriangleup} {2}{mathb}{"98}% name to be checked
\DeclareMathSymbol{\smalltriangledown} {2}{mathb}{"99}% name to be checked
\DeclareMathSymbol{\smalltriangleleft} {2}{mathb}{"9A}% name to be checked
\DeclareMathSymbol{\smalltriangleright}{2}{mathb}{"9B}% name to be checked
\DeclareMathSymbol{\blacktriangleup} {2}{mathb}{"9C}% name to be checked
\DeclareMathSymbol{\blacktriangledown} {2}{mathb}{"9D}% name to be checked
\DeclareMathSymbol{\blacktriangleleft} {2}{mathb}{"9E}% name to be checked
\DeclareMathSymbol{\blacktriangleright}{2}{mathb}{"9F}% name to be checked

\usepackage{xspace} % insert space when needed.
\newcommand{\xps}{ExPoSe\xspace}
\renewcommand{\xps}{WikiCat\xspace}
\newcommand{\expose}{ExPoSe WikiCat Browser\xspace}
\renewcommand{\expose}{WikiCat Browser\xspace}
\newcommand{\spn}{search\:-\:powered navigation\xspace}

\newcommand{\maingoal}{to investigate information seeking behavior and user experience when navigation is empowered by a search functionality.}

\newcommand{\rqone}{What is the effect of search powered navigation on user behavior in different types of exploratory search tasks?}

\newcommand{\rqtwo}{Does empowering navigation with search improve user experience in different types of exploratory search tasks?}

\pgfplotstableread{
j	BT	BT-spn	FT	P-spn
1	4.4	13.3	10.2	9.3
2	8	17.3	8.1	7.3
3	9.1	15.5	32	19
4	42.9	10.1	14	12.1
5	31.2	40.9	34	51.4
6	2.3	1.6	1.2	0.6
7	2.1	1.3	0.5	0.3
}\tableone

\pgfplotstableread{
l	BT	BT-spn	FT	P-spn
1	6	6	5	5
2	30	17	9	6
3	38	20	25	18
4	14	33	27	28
5	9	14	23	29
6	3	10	11	14
}\tabletwo

\pgfplotstableread{
r	BT-Clicks	BT-Scores	FT-Clicks	FT-Scores
1	39	34	38	77
2	14	26	16	10.7
3	10	14	11	6
4	7	10	10	3
5	9	9	7	2.1
6	21	7	19	1.2
}\tablethree

\copyrightyear{2017}
\acmYear{2017}
\setcopyright{acmlicensed}
\acmConference{ACM SIGIR International Conference on the Theory of Information Retrieval}{}{October 1--4, 2017, Amsterdam, The Netherlands}
\acmPrice{15.00}
\acmDOI{https://doi.org/10.1145/3121050.3121105}
\acmISBN{ISBN 978-1-4503-4490-6/17/10}

\begin{document}

	\title{Search Versus Navigation: \\ Which Deserves to Go in Exploratory Search?}
    \title{Does Search Help Navigation in Exploratory Search?}
    \title{Search and Navigation,Two Aspects of Exploratory Search}

	\title{Search and Navigation, Two Ingredients of Exploratory Search}
	\title{Search and Navigation, Two Ingredients of Exploratory Search}
	\title{Search Powered Navigation}
	\title{On Search Powered Navigation} %???

\author{Mostafa Dehghani}
\affiliation{%
  \institution{University of Amsterdam}
  %\streetaddress{}
  %\city{} 
%   \country{The Netherlands}
}
\email{dehghani@uva.nl}

\author{Glorianna Jagfeld}
\affiliation{%
  \institution{University of Stuttgart}
  %\streetaddress{}
  %\city{} 
%   \country{Germany}
  }
\email{jagfelga@ims.uni-stuttgart.de}

\author{Hosein Azarbonyad}
\affiliation{%
  \institution{University of Amsterdam}
  %\streetaddress{}
  %\city{} 
%   \country{The Netherlands}
}
\email{h.azarbonyad@uva.nl}

\author{Alex Olieman}
\affiliation{%
  \institution{University of Amsterdam}
  %\streetaddress{}
  %\city{} 
%   \country{The Netherlands}
}
\email{olieman@uva.nl}
  
\author{Jaap Kamps}
\affiliation{%
  \institution{University of Amsterdam}
  %\streetaddress{}
  %\city{} 
%   \country{The Netherlands}
}
\email{kamps@uva.nl}

\author{Maarten Marx}
\affiliation{%
  \institution{University of Amsterdam}
  %\streetaddress{}
  %\city{} 
%   \country{The Netherlands}
}
\email{maartenmarx@uva.nl}



% We no longer use \terms command

\begin{abstract}
        Query-based searching and browsing-based navigation are the two main components of exploratory search.  Search lets users dig in deep by controlling their actions to focus on and find just the information they need, whereas navigation helps them to get an overview to decide which content is most important.
		In this paper, we introduce the concept of \emph{search powered navigation} and investigate the effect of empowering navigation with search functionality on information seeking behavior of users and their experience by conducting a user study on exploratory search tasks, differentiated by different types of information needs.
		%, i.e. focused\:-\:topics and broad\:-\:topics.
		%
		Our main findings are as follows:
		First, we observe radically different search tactics.  Using search, users are able to control and augment their search focus, hence they explore the data in a depth\:-\:first, bottom\:-\:up manner.
		Conversely, using pure navigation they tend to check different options to be able to decide on their path into the data, which corresponds to a breadth\:-\:first, top\:-\:down exploration. 
		Second, we observe a general natural tendency to combine aspects of search and navigation, however, our experiments show that the search functionality is essential to solve exploratory search tasks that require finding documents related to a narrow domain. %on focused topics.
        Third, we observe a natural need for search powered navigation: users using a system without search functionality find creative ways to mimic searching using navigation.
        %\vskip -0.5ex
        %\small  % Changes line spacing, not font -- throughout...
%\mypar{KEYWORDS} ~~~ Search, Navigation, Exploratory Search
        %\vskip -1ex
\end{abstract}
%\keywords{Exploratory Search, Search, Navigation}
\maketitle

%A category including the fourth, optional field follows...
%\category{D.2.8}{Software Engineering}{Metrics}[complexity measures, performance measures]
%% A category with the (minimum) three required fields
%%\category{H.4}{Information Systems Applications}{Miscellaneous}
%\category{H.3.3}{Information Storage and Retrieval}{Information Search and Retrieval}[Query formulation, Relevance feedback, Search process]
%%H. Information Systems
%%H.3 INFORMATION STORAGE AND RETRIEVAL
%%H.3.3 Information Search and Retrieval
%%Clustering
%%Information filtering NEW!
%%Query formulation
%%Relevance feedback NEW!
%%Retrieval models
%%Search process
%%Selection process
%
% I. Computing Methodologies
% I.2 ARTIFICIAL INTELLIGENCE
% I.2.6 Learning (K.3.2)
% Concept learning
% I.2.7 Natural Language Processing
% Language models
% 
% Algorithms, Design, Documentation, Economics, Experimentation, Human Factors, Languages, Legal Aspects, Management, Measurement, Performance, Reliability, Security, Standardization, Theory, Verification
%\terms{Theory, Experimentation}
%
%\keywords{}  % NOT required for Proceedings

% JK Smaller version of the above:

% \vspace{1mm}
% \noindent
% {\bf Categories and Subject Descriptors:} %
% {I.2.6} {[\textbf{Artificial Intelligence}]: Learning}\:---\:\emph{Concept learning}; {I.2.7} {[\textbf{Artificial Intelligence}]: Natural Language Processing}\:---\:\emph{Language models}

%--------------------------------------------------------------------
\shrink %\sshrink
\section{Introduction}	
	Knowledge graphs and other hierarchical domain ontologies hold great promise for complex information seeking tasks, yet their massive size defies the standard and effective way smaller hierarchies are used as a static navigation structure in faceted search or standard website navigation.  As a result, we see only limited use of knowledge bases in entity surfacing for navigational queries, and fail to realize their full potential to empower search.
	%
	%However, providing support to users in all stages of information acquiring is required. This spans the complete process from the initial formulation of the area of interest to the discovery of relevant sources and the establishment of relations among them
	%
	%to be supported by going beyond one-time interactions and allowing users to progressively seek for information and most of the time, the process contain progressive discovery based on the available information, including the actions of lookup, browsing, analysis, and exploration.
	%
	%To support such information seeking behaviour, providing support to users in all stages of information acquiring is required. This spans the complete process from the initial formulation of the area of interest to the discovery of relevant sources and the establishment of relations among them~\citep{White:2007}.
	Seeking information in structured environments consists of two main activities: \emph{exploratory browsing} and \emph{focused searching}~\citep{White:2009, Wilson:2010,Dehghani:2017}.
	Exploratory browsing refers to activities aimed at better defining the information need and increasing the level of understanding of the information space, while focused searching includes activities such as query refining and comparison of results, which are performed after the information need has been made more concrete.
	Based on the interplay of these two actions, a search system is supposed to provide a connected space of information for the users to \emph{navigate}, as well as \emph{search} to adjust the focus of their browsing towards useful content. 
	
	%So, in order to design an effective exploratory search system, it is important to understand the extent to which browsing behaviours differ in terms of the interaction flow and targeted information.
	%related work
	In this paper, we introduce the concept of \emph{Search Powered Navigation (SPN)}, which enables users to combine navigation with query based searching in a structured information space, and offers a way to find a balance between exploration and exploitation.  We hypothesize that SPN enables users to exploit the semantic structure of a large knowledge base in an effective way.  	
	We test this hypothesis by conducting a user study in which users are  engaged in exploratory search activities and investigate the effect of SPN on the variability in users' behaviour and experience. We employed an exploratory search system on parliamentary data in two modes, \emph{pure navigation} and \emph{search powered navigation}, and tested two types of tasks, broad\:-\: and focused\:-\:topic tasks.

	In our study, the primary goal is \emph{\maingoal}\ 
	We break it down into the following two research questions:
	\begin{description}
		  \setlength{\itemsep}{2pt}
		  \setlength{\parskip}{-0.5pt}%
		  \setlength{\parsep}{-0.5pt}
		\item[\textbf{RQ1}] \rqone
		\item[\textbf{RQ2}] \rqtwo
		%\item[\textbf{RQ3}] \rqthree
	\end{description}
	\begin{figure*}[t]
\makebox[\linewidth][c]{
\centering
\begin{subfigure}[t]{0.32\linewidth}
\centering
\includegraphics[width=\linewidth]{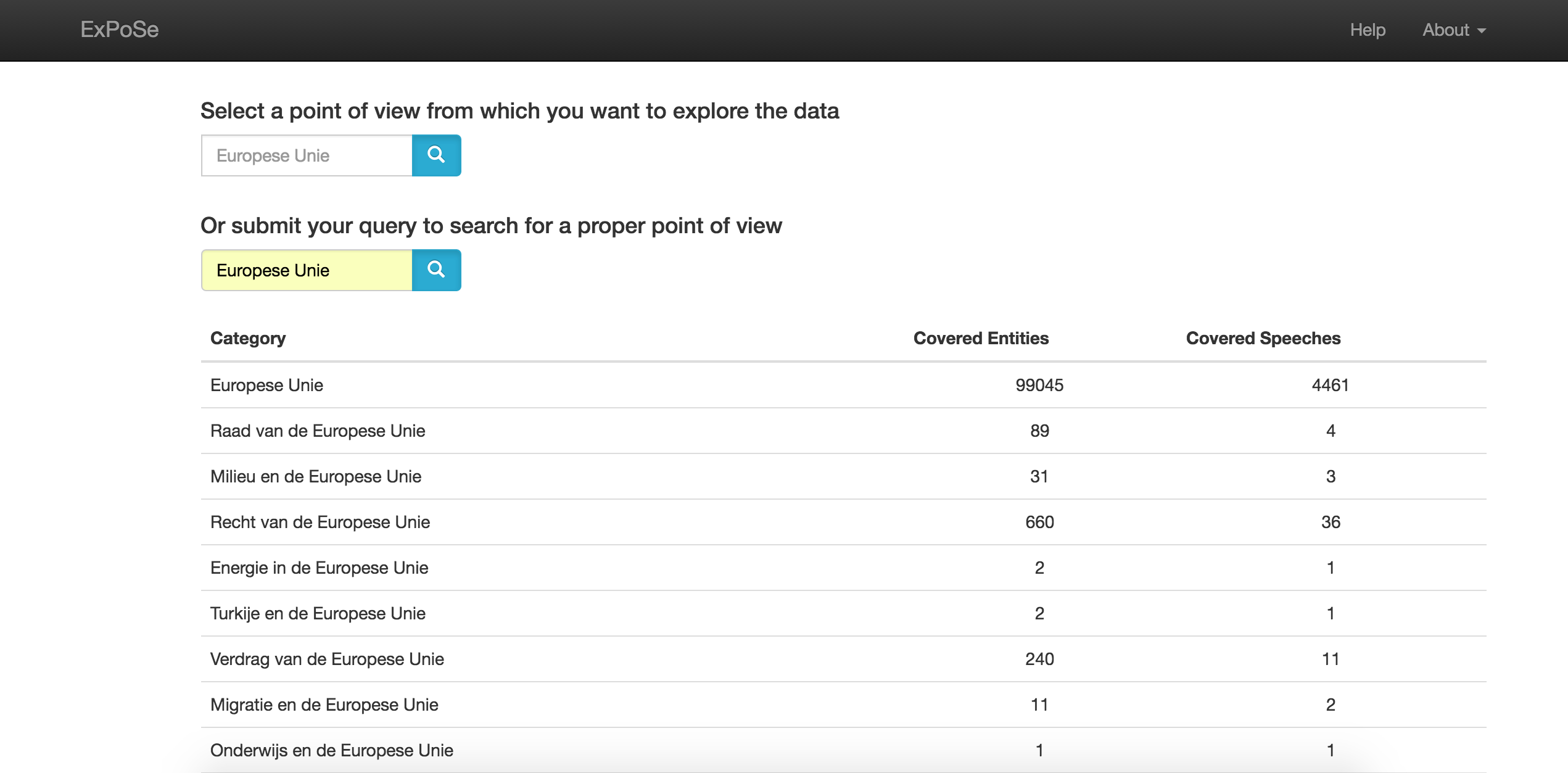}
\caption{\label{fig:expose1}\footnotesize{Starting page of the \expose system, where the users can choose a Wikipedia category as the point of view from which to explore the data.}}
\end{subfigure}
\hfill
\begin{subfigure}[t]{0.32\linewidth}
\centering
\includegraphics[width=\linewidth]{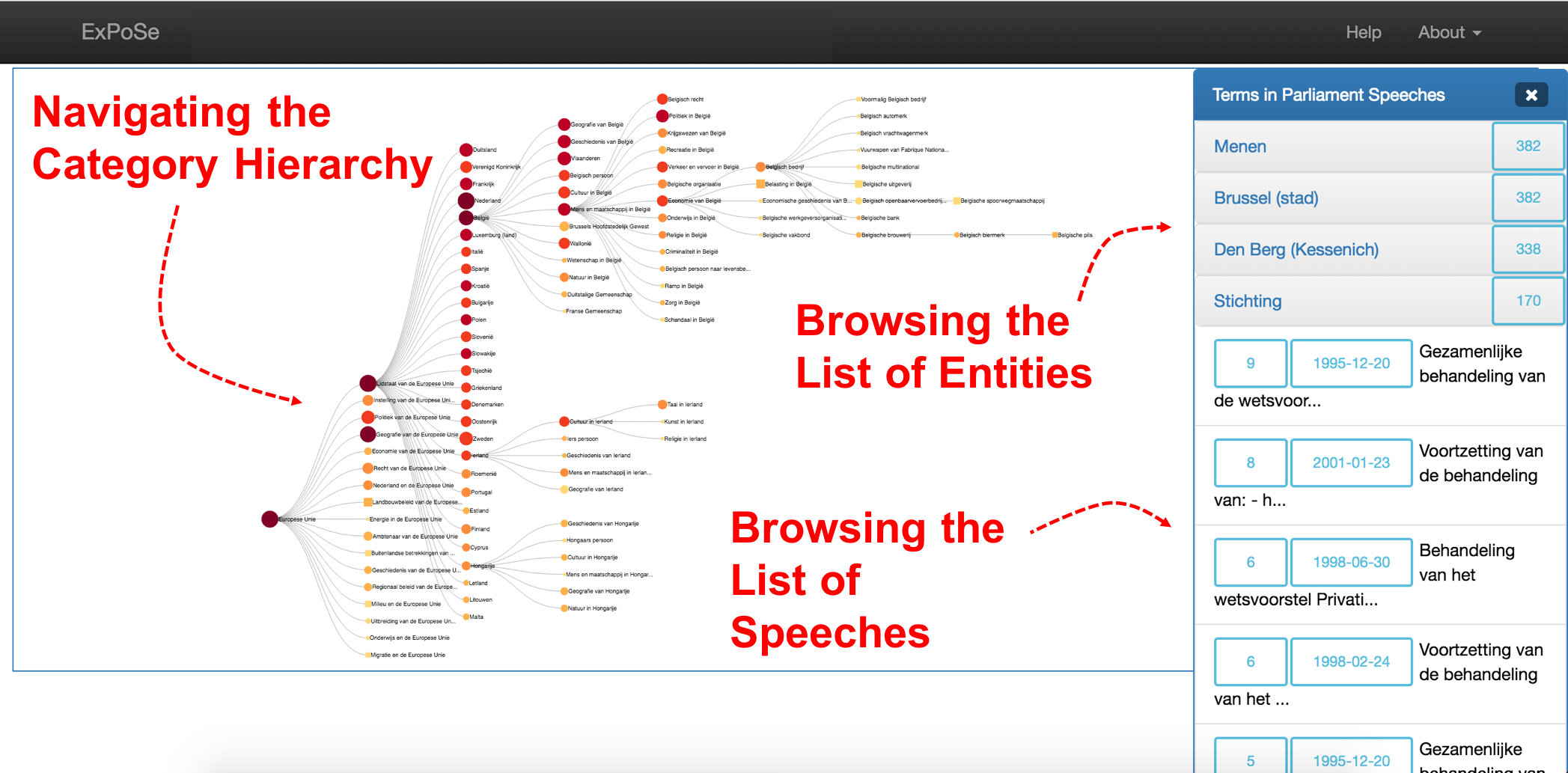}
\caption{\label{fig:expose2}\footnotesize{Sample expanded hierarchy. The right panel provides the list of entities and speeches related to the selected category.}}
\end{subfigure}
\hfill
\begin{subfigure}[t]{0.32\linewidth}
\centering
\includegraphics[width=\linewidth]{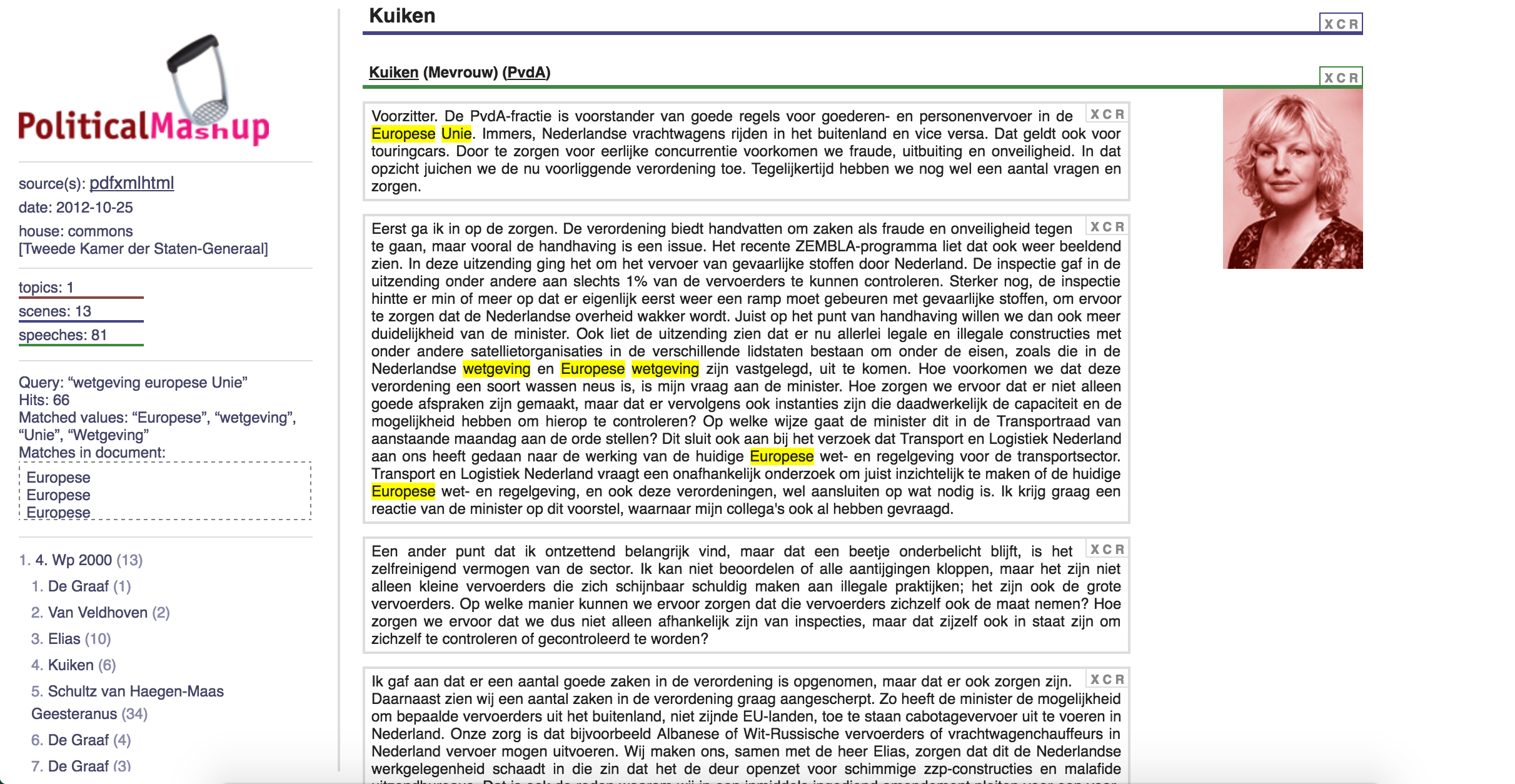}
\caption{\label{fig:expose3}\footnotesize{Sample browsed speech. It contains information about the debate and the content of all its speeches. The source entity is highlighted in the text.}}
\end{subfigure}
}
\vspace{-10pt}
\caption{\expose system}
\vspace{-18pt}
\end{figure*}

%
%\hfill
%

\begin{figure}[t]
\vspace{-3pt}
\centering
\includegraphics[width=0.9\linewidth]{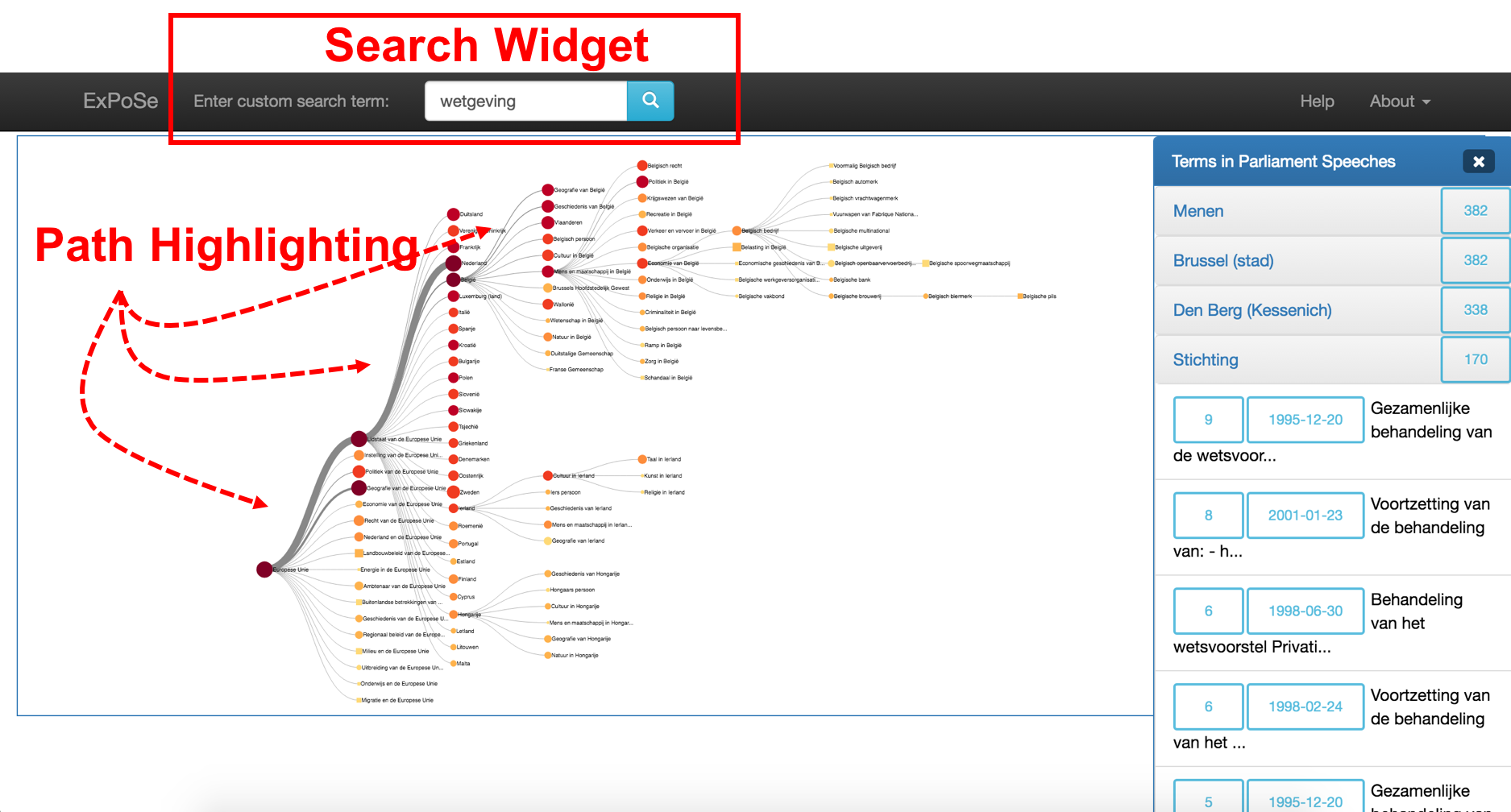}
\vspace{-5pt}
\caption{\label{fig:expose4}\footnotesize{The \expose system with SPN. The users can submit a textual query which highlights the paths in the hierarchy based on their likelihood of leading to speeches relevant to the query.}}
\vspace{-15pt}
\end{figure}	
    There is a large body of research focusing on user behavior in exploratory search from different angles.
	\citet{Athukorala:2014:CIKM} investigated user behavior in terms of narrowing or broadening queries in exploratory search. 
    \citet{Diriye:2010} discussed different factors that generally influence exploratory search like objective, search activities, conceptual complexity, procedural complexity, and domain knowledge.
	%
	%Changes in the user browsing behaviour during complex search tasks is also investigated by~\citep{White:2009}.
	%
	In terms of investigating user behavior during search, there are also relevant studies conducted in the context of general web search~\citep{HsiehYee:2001} and website browsing~\citep{Blackmon:2005}. \citet{White:2007} also studied the interaction flow in web search and investigated user behavioural variability based on users' queries,interactions, time, and the types of visited webpages.
	
	% Moved to end...
	%%Overall, we find that search is an important means for the users, especially in focused\:-\:topic tasks. However, the presence of a search function does not prevent the users from exploring a significant portion of the data.	
	%Search lets them control their own actions and find just the information they are looking for, while navigation guide them where they should click and decides on their behalf which content is more important than other.
	%
	%Our results can be directly applied to the design of tools to support effective exploratory search. 
	
\shrink 
\section{Case Study System Overview}
	\label{sec:expose}
%	Parliamentary proceedings, and political conversational data in general, capture and reflect many significant events in the world. While is important to be able to search and explore this kind of data, they are characterised by a complex narrative structure, which makes them difficult to access for many people.
	%
% 	Using structured data to organize unstructured information is one of the most common approaches for supporting complex search tasks, including exploratory search~\citep{Marie:2014}. Furthermore, exploring concepts and their relations are the major characteristics of this type of search~\citep{Vakkari:2010}.
	For our study, we have used an exploratory search system, the \expose, which maps parliamentary data into a conceptually structured space to facilitate exploring the data.

	\sshrink 
	\mypar{\expose.}
	The \expose is a search system developed for exploratory search to browse and investigate parliamentary debates from a particular point of view.
	It projects the parliamentary speeches to the Wikipedia categorical structure, based on Wikipedia named entities mentioned in the parliamentary speeches. 
    A customized entity linker for parliamentary conversation ~\citep{Olieman:2014,Dehghani:2017} is employed to recognize and disambiguate both the general and domain specific entities from all the speeches.
    
	In this system, first, the user has to chose the point of view from the set of Wikipedia categories.
	To assist users in choosing a good staring point of view, the \expose
	provides a search tool,	which was first introduced to the system for the experiments presented in this paper.
	Given a free text query, the search tool retrieves a ranked list of categories according to the BM25 similarity of the query to the content of the debates grouped under each category (Figure~\ref{fig:expose1}). In the selected hierarchy, each node is a Wikipedia category representing one topic.
    The users can explore the category hierarchy by clicking on a node which expands its descendants, i.e. sub-categories (Figure~\ref{fig:expose2}). Importance and recency of the categories are visualized by the size and color of the nodes, respectively.
    The importance of each node denotes how much the topic of that category is addressed in the parliamentary debates, based on the frequency of entities under this category.
    The recency of each node shows how recently the topic of the category was discussed, calculated from the dates of the debates in this category.
    
    %Nodes can be expanded to display the next level of the hierarchy.
    
    Besides expanding each category, the users can browse the list of entities under the category by clicking on the category name.
    For each entity, the users can browse the list of all speeches in which this entity occurs.
    The list of speeches is ordered by the number of occurrences of the corresponding entity.
    For each speech the title of the debate and its date is shown.
    By clicking on the speech title, the full content of the speeches with the related entities highlighted in the text is displayed (Figure~\ref{fig:expose3}).
	
	As example of the type of search questions that can be addressed with the \expose constitutes analysing the relation between national laws of the European country and the European union legislation. 
	As a reasonable approach, we would like to investigate when EU legislation was discussed in the Dutch parliament and in the context of which (proposed) Dutch laws, i.e. we want to project debates in the Dutch parliament to topics related to the European Parliament in terms of both the subject matter and time.
\newcommand{\multilinecell}[2][c]{%
\begin{tabular}[#1]{@{}c@{}}#2\end{tabular}
}

\begin{table*}[tbp]
\centering
\caption{\label{tbl:tbl1} Per-session statistics from search sessions for both systems in both types of tasks averaged over all search sessions. Numbers marked by $^\blacktriangleup$ or $^\blacktriangledown$ differ significantly from all other numbers in the same column (two-tailed t-test, p-value $<$ 0.05).}
\vspace{-10pt}
%\makebox[\linewidth][c]{
{\renewcommand{\arraystretch}{1.35} 
\begin{adjustbox}{max width=0.85\textwidth}
\begin{tabular} 
{l c c c c c c c c c}
%{@{}c@{~~}l@{~~~}c@{~~}c@{~~}c@{~~}c@{~~}c@{~~}c@{~~}c@{}c@{}}
\toprule
%\fontsize{6}{7}\selectfont
\textbf{Task} & \textbf{System} & 
\multilinecell{\textbf{avg. num. of} \\ \textbf{root} \\ \textbf{selection}} & 
\multilinecell{\textbf{avg. depth of} \\ \textbf{selected root in} \\ \textbf{Wikipedia hierarchy}} & 
\multilinecell{\textbf{avg. num. of} \\ \textbf{node} \\ \textbf{expansions}} &
\multilinecell{\textbf{avg. num. of} \\ \textbf{entity list} \\ \textbf{loading}} & \multilinecell{\textbf{avg. num. of} \\ \textbf{speech list} \\ \textbf{loading}} &  \multilinecell{\textbf{avg. num. of} \\ \textbf{speech content} \\ \textbf{browsing}} & \multilinecell{\textbf{avg. session} \\ \textbf{duration} \\ \textbf{(min)}} & \multilinecell{\textbf{avg. rating} \\ \textbf{summaries/speeches}\\ \textbf{(10 point scale)}}\\
\midrule
\multirow{2}{*}{\rotatebox{90}{\multilinecell{Broad\\Topic}}} & \xps & \Large{3.1 } & \Large{5.1 } & \Large{15.1 } & \Large{29.5 } & \Large{16.7 } & \Large{18.2 } & \Large{22.2 } & \Large{5.12 }
\\[2pt]
%\cmidrule(lr){2-10}
& \xps + SPN & \Large{2.2 } & \Large{4.7 } & \Large{11.8 } & \Large{20.8 } & \Large{12.4 } & \Large{13.1 } & \Large{18.2 } & \Large{7.78 } \\
\midrule
\multirow{2}{*}{\rotatebox{90}{\multilinecell{Focused\\Topic}}} & \xps & \Large{ 8.7$^\blacktriangleup$ } & \Large{ 9.3$^\blacktriangleup$ } & \Large{ 9.5$^\blacktriangledown$} & \Large{ 31.6$^\blacktriangleup$ } & \Large{ 9.2 } & \Large{ 4.9 } & \Large{ 33.4 } & \Large{3.44 } \\[2pt]
%\cmidrule(lr){2-10}
& \xps + SPN & \Large{ 3.3 } & \Large{ 6.3 } & \Large{ 13.9 } & \Large{ 16.4 } & \Large{ 7.6 } & \Large{ 4.1 } & \Large{ 27.1 } & \Large{ 4.62 } \\
\bottomrule 
\end{tabular}
\end{adjustbox}
%}
}
\vspace{-10pt}
\end{table*}
\definecolor{b}{HTML}{4981CE}
\definecolor{g}{HTML}{859C27}
\definecolor{r}{HTML}{B22222}
\definecolor{o}{HTML}{FF6600}

\begin{figure*}[t]
% \makebox[\linewidth][c]{
% \centering
\centerline{
\begin{minipage}[b]{0.6\textwidth}
\centering
\includegraphics[width=0.99\linewidth]{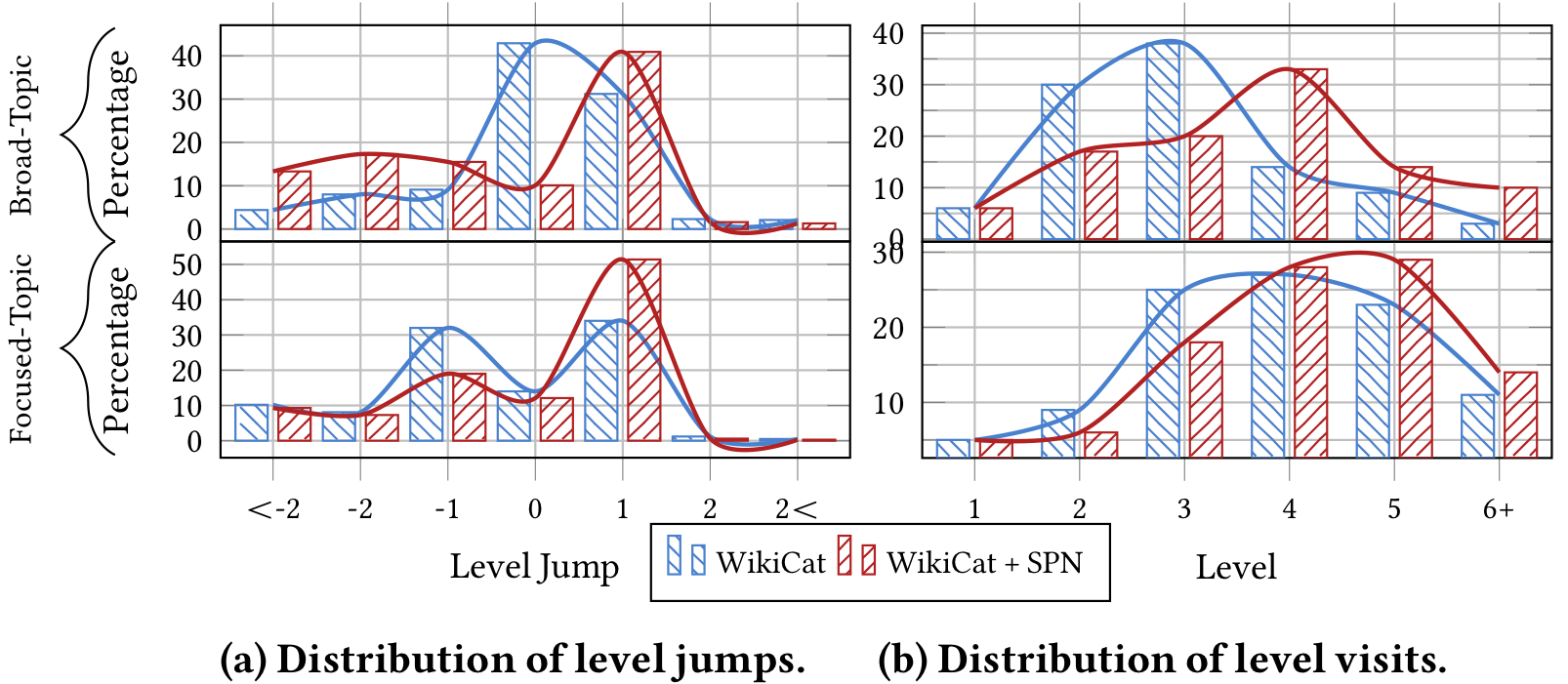}
\vspace{-10pt}
\caption{Distribution of level jumps and visits for focused and broad topics.}
\end{minipage}
\hspace{\stretch{2}}
\begin{minipage}[b]{0.38\textwidth}
\centering
    \centering
\includegraphics[width=0.85\linewidth]{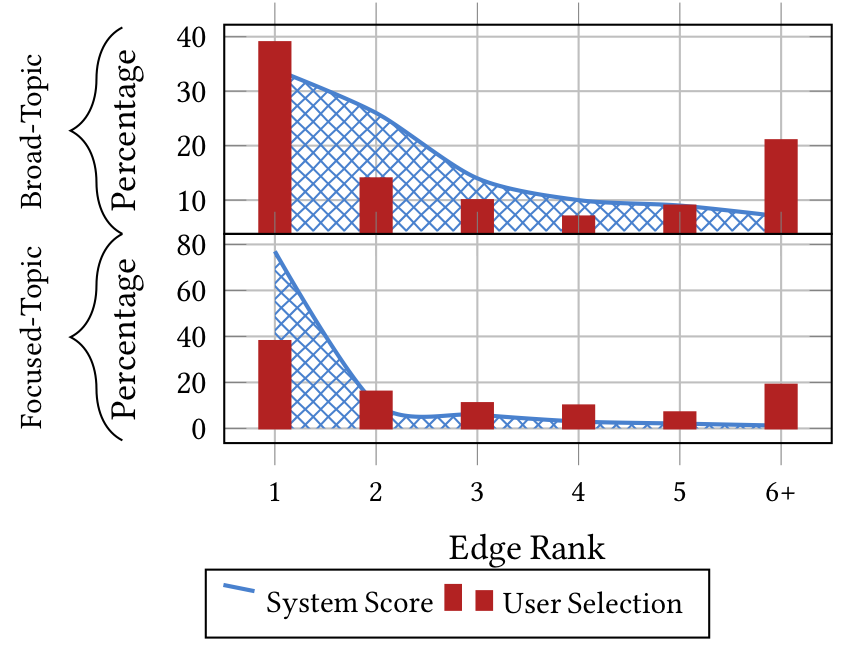}
    \vspace{-10pt}
    \captionof{figure}{\label{chart:rank}\small{Percentage of selected edges on different ranks versus distribution of scores assigned to edges.}}
\end{minipage}
}
\vspace{-15pt}
\end{figure*}
    \sshrink
    \mypar{Search\:-\:Powered Navigation.}	
	In order to investigate the effect of search on the browsing behaviour of the users, we designed \spn (SPN) for the \expose. SPN provides the ability to \emph{search} during the data navigation.
	It allows the users to submit a textual query at any step in the navigation.
	The system ranks different navigation paths based on their likelihood to leading to data relevant for the query~(Figure~\ref{fig:expose4}).
	
	SPN is integrated as a search bar in the navigation window of the \expose.
	Every time the user submits a query, the system calculates the similarity of the full content of all speeches in the collection with the given query.
	The scores at the speech level are used to calculate the scores at the entity level based on their occurrences in different speeches and their corresponding scores. 
	Finally, the normalized scores of all entities under each category are summed to obtain the likelihood score of the category.
	This score determines the weight of the edge from the previous level of the hierarchy to the category. 
	
	In the user interface, the weights of different edges are represented by their thickness.
	At each level in the navigation hierarchy, SPN provides a ranking of paths that are likely to lead to speeches relevant for the query.
	In line with the general purpose of the \expose system, which is exploratory search, SPN does not provide a direct ranking of speeches but hints the user to more promising paths.
	In that way, the users still can explore the full search space but obtain some guidance in their decision making process.
	In our experiments we show that the imposed bias by SPN for steering users to relevant information does not keep them from exploring the space.
    
	\shrink \sshrink
	\section{Experimental Setup}
	\label{sec:exp}
    We have conducted a descriptive user study~\citep{Kelly:2009} by asking the users to do exploratory search tasks on Dutch parliamentary debates from 1994 to 2014, once by means of the pure navigational system and once using the system with SPN.
	
	With regard to the requirements of simulated work task situations~\citep{Borlund:2003}, we selected 14 search tasks, comprising seven broad\:-\:topic and seven focused\:-\:topic tasks and defined task scenarios for the users.
	Broad\:-\:topic tasks, on the one hand, are supposed to be addressed by a diverse set of debates and there is no concrete answer to them.
	Focused\:-\:tasks, on the other hand, are supposed to be addressed by a small fraction of debates. 
	%
	%TODO are we allowed to submit an appendix? then we could put the list of tasks there
	In the broad\:-\:topic tasks, the users had to find out the general attitude or opinion of the Dutch parliament on the following seven topics: \emph{Immigration}, \emph{Islam}, \emph{World War II}, \emph{Taxes}, \emph{Holocaust}, \emph{Dutch Golden Age}, \emph{European Union}. 
	In the focused\:-\: topic tasks, the users should find Dutch parliamentary debates related to the following seven particular cases: 
	\emph{Unsuccessful operation of ``baby Jelmer'' in Groningen University Hospital}; 
	\emph{The trial and death of Slobodan Milosevic in the Hague}; 
	\emph{Role of Dutchbat soldiers in the Srebrenica massacre};
	\emph{New year's fire in a cafe in Volendam}; 
	\emph{The fireworks disaster in Enschede}; 
	\emph{H5N1 (a variant of the bird flu virus)}; and 
	\emph{Fitna (an anti\:-\:Islam film by the Dutch parliamentarian Geert Wilders)}.
	
	In our study, 14 participants (5 undergraduate and 9 graduate students) were enrolled, six male and eight female, who had an average age of 26 (SD = 2.01). All participants had more than 10 years of computer experience, a fairly high ability of online search, but inexperienced in the political domain and Dutch parliament. 	
	The participants were each given two broad\:-\:topic and two focused\:-\:topic tasks.
	They had to complete one broad\:-\:topic task and one focused\:-\:topic task using the original \expose system, and the two remaining tasks using the system with SPN.
	To control the order bias, we randomly divided the users into two equally sized groups.
	The first group started off with the baseline system, while the second group used the system with SPN first.
	In all experiments, the participants were asked to provide a set of relevant speeches along with a short summary of the obtained information.
	
	%Prior to the experiments, the participants were trained to work with the systems. 
	Based on our setup, each task was performed twice with the original \expose system and twice with the system with SPN, each time by a different participant.
	In total, we had 56 search sessions. 
	Additionally, the users were asked to fill out a basic pre-search questionnaire with their general information, including age, education, gender, and a self-assessment of their domain knowledge and their ability to use search systems.
	After each task, they filled out a post-search questionnaire to rate the difficulty of the task and their satisfaction~\citep{Kelly:2015}.
	During the experiments, we logged all interactions of the users with systems. 

\shrink 
\section{Results and Discussion}
\label{sec:res}
	In this section, we analyze the effect of our different conditions, search vs. no search, broad vs. focused topic tasks on the navigational behavior of the users and their experience during search. 
	We describe the navigational model in terms of \emph{level visits} and \emph{level jumps}.
	A \emph{level visit} is counted if the user expands the children of a node on a certain level or loads the list of entities of the node.The \emph{level jump} denotes the difference of the node levels of a pair of consecutively visited nodes.
	Additionally, we recorded statistics on the different actions the users can take during browsing which implicitly indicate how user experience changes in terms of time and effort. The results are shown in~Table~\ref{tbl:tbl1}.

\shrink
	\subsection{Effect of SPN on user behaviour} 
	We analyze in what ways adding search functionality influences the navigational behavior.
	According to the plots in the upper part of Figure~3a, for broad\:-\:topic tasks the users of the \expose system tend to visit nodes from the same layer when exploring the space (high percentage of zero level jumps), which corresponds to a \textbf{breadth-first} navigation.
	However, as the lower part of Figure~3a shows, when their navigation is empowered by a search utility, they tend to have more forward moves and visit nodes in the next levels (high percentage of level jumps $=1$), which corresponds to a \textbf{depth-first} exploration of the hierarchy.
	For the focused\:-\:topic tasks, the traversal approach is \textbf{depth-first} for both systems.
	This pattern can be explained by the information need.
	Since the users are not able to evaluate the quality of a taken path until they are deep enough to access a narrow part of the search space.
	The main difference in the browsing behaviour of the users is that they backtrack a lot using the baseline system (high percentage of negative level jumps).
	The reason for this is that the users need to try more different paths when they have no strong clue of the content of the documents in the deeper levels.
	Figure~\ref{chart:rank} demonstrates the percentage of the selected edges on different ranks versus the distribution of their scores. %assigned to edges over different ranks
	%in both tasks
	In the broad\:-\:topic tasks, the system assigns rather high scores to the edges in the second and third ranks, while in the focused\:-\:topic tasks, the score distribution is pretty skewed, where the top-ranked edge receives a considerable share of the score distribution.
	Regardless of the task, the users choose the top-ranked edge at each point to deepen their exploration in about 40\% of the cases. %, the edge in the first rank is chosen to be pursued. 
	Interestingly, the users were still motivated to explore lower ranked edges irrespective of the score distribution of the edges.
	Moreover, contrary to the concern that search may limit the extent of exploration, the users are more confident to go deeper in the hierarchy with the system with SPN (cf. Figure~3b).
	%As the users reported in the post-search questionnaire, the suggestions from the search utility decreased their risk of taking an unhelpful navigation path.
	
\shrink
	\subsection{Does SPN improve user experience?} 
	First we investigate the general user behavior with regards to the task type, then we study how empowering navigation with search improves user experience in each type of task.

	We look at the effect of different tasks on the navigational model.
	As expected the users explore more of the search space in the broad\:-\:topic tasks because they require more browsing.
	In average the rate of node expansions and data loading including entities, speeches, and content of speeches displayed in Table~\ref{tbl:tbl1} are higher in both systems compared to %their corresponding systems in
	the focused\:-\:topic tasks.
	As can be seen from Figure~3b, in general the users tend to visit deeper levels in focused\:-\:topic tasks. %compared to the broad\:-\:topic tasks. 
	It is noteworthy that in both systems, all the data of the deeper levels is also accessible from higher levels. However, with regard to the sorting strategy of displayed information, the more popular the data is (i.e. more frequent entities in each category), the higher it is ranked in the browsing panel.
	Hence, for the broad\:-\:topic tasks, which are mostly addressable with popular entities and speeches,
	the users find their needed information on top of the lists already at higher levels and they see no need to go deeper.
	In contrast, for the focused\:-\:topic tasks, which are mostly related to particular entities that are not necessarily popular overall, in the higher levels the users need to scroll down a long list to find a related data among lots of unrelated data.
	Thus, according to our observations, they prefer to descend in the hierarchy to narrow down the search space.
	
    We further investigated whether the search functionality is a useful addition to an exploratory search system.
	As can be seen in Table~\ref{tbl:tbl1}, the average number of loaded entities, speeches or their content is generally lower for the system with SPN compared to the original \expose system in both types of tasks.
	This can be attributed to the fact that users skip loading the intermediary data when their navigation is facilitated by search.
	They quickly get to the target content, which leads to shorter search sessions. 
	We assessed the quality of the search results by asking two experts in the Dutch political domain to rate set of relevant speeches and summaries retrieved by the users.
	The assessors were not informed about the participants and the system used to compile the summaries and speeches.
	As can be seen from the last column in Table~\ref{tbl:tbl1}, the users were more successful in the broad\:-\:topic tasks
	compared to the focused\:-\:topic tasks, regardless of the search system.
	However, the search functionality improved the quality of the results in both tasks significantly.
	In summary, we conclude that the search lays the ground for a better navigation of the users, enabling them to explore better content.	
	
	In our study, in the focused\:-\:topic tasks, the users are supposed to find a needle in a haystack of data. 
	Using the original \expose system, 
	%they are able to browse the content based on the relations between entities and Wikipedia categories.Hence, 
	if they fail to express the information need using a set of proper entities, it is less likely that they find paths leading to the relevant content.
	%In contrast, by using the search utility, they might be able to articulate their information need using simple textual queries that do not necessarily contain entities.
	%We were particularly interested how the users' behaviour changes when they struggle to find the information in these situations.
	%
	The statistics in Table~\ref{tbl:tbl1} reveal an interesting behavior of the users' faced with the original \expose system to perform focused\:-\:topic tasks.
	In this setting, the average number of node expansions is significantly lower compared performing the same type of tasks with the system with SPN. 
	This runs contrary to our expectations, since in the system with no navigation aid, users are supposed to try different paths by expanding a lot of nodes, as it is the case in the broad\:-\:topic tasks.
	However, considering the high average number of root selections and their high average depths %of the selected roots
	in the focused\:-\:topic tasks using the original \expose system, we noticed that the users seemed to creatively simulate the search in their navigation when they needed it.
	They went back to the starting page of the system, where they can choose the root of the hierarchy they want to navigate through, using the search functionality provided for choosing the root node.
	Then, instead of navigating from there to deep levels, they shallowly evaluate the selected hierarchy and again go back to the starting page to choose a better root node by reformulating their query.
	They repeat this procedure until they get access to a proper root node which is fairly deep in the hierarchy.
	Thus, the users  try different nodes as the starting node using the search instead of trying to locate them via navigation. 

	%---------------------------------------
% 	\sshrink
	\section{Conclusions}
	\label{sec:con}
	%---------------------------------------
	%
	%The main aim of this paper was \textsl{\maingoal}\
    We introduced the concept of ``search powered navigation'' as a way to empower users with the semantic structure of large knowledge bases, and build a proof-of-concept system to investigate the viability of this approach to support exploratory search tasks.
    In a user study, we showed that a search functionality can be a viable and effective addition to a navigational exploratory search system. This was reflected in faster task completion and better search results using the system with the additional search feature.
    While users tended to search the data in a breadth-first manner in the pure navigational baseline exploratory search system, an additional search functionality lead to a depth-first investigation of the data. 
    %TODO removed due to space constraints
    %In the contrary, the search functionality encouraged the users to access deeper levels of the hierarchy and thus broadening their exploration of the data.
    Furthermore, we found that a search functionality is essential to be able to efficiently address focused\:-\:topic tasks.
	%
%	In future work, we plan to extend our study with more users and more dimensions.
%	For instance, we would like to take personality features of the users into account and compare domain experts to novice users. %, and collect more explicit feedback from the users by means of a more detailed post-search questionnaire.
	%We will analyse this information along with the log-based data. 
	% 
	%Overall, our results inform the design of tools to support effective exploratory search in the context of large scale knowledge bases.

% \sshrink
%ACKNOWLEDGMENTS are optional
%\section*{Acknowledgments}
\mypar{Acknowledgments}  \small This research is funded in part by the Netherlands Organization for Scientific Research (NWO;  ExPoSe project, NWO CI \# 314.99.108).

% \shrink
%\bibliographystyle{ACM-Reference-Format}
\bibliographystyle{abbrvnat}
\bibliography{ref}

\begin{thebibliography}{12}
\providecommand{\natexlab}[1]{#1}
\providecommand{\url}[1]{\texttt{#1}}
\expandafter\ifx\csname urlstyle\endcsname\relax
  \providecommand{\doi}[1]{doi: #1}\else
  \providecommand{\doi}{doi: \begingroup \urlstyle{rm}\Url}\fi

\bibitem[Athukorala et~al.(2014)Athukorala, Oulasvirta, Glowacka, Vreeken, and
  Jacucci]{Athukorala:2014:CIKM}
K.~Athukorala, A.~Oulasvirta, D.~Glowacka, J.~Vreeken, and G.~Jacucci.
\newblock Narrow or broad?: Estimating subjective specificity in exploratory
  search.
\newblock In \emph{CIKM '14}, pages 819--828, 2014.

\bibitem[Blackmon et~al.(2005)Blackmon, Kitajima, and Polson]{Blackmon:2005}
M.~H. Blackmon, M.~Kitajima, and P.~G. Polson.
\newblock Tool for accurately predicting website navigation problems,
  non-problems, problem severity, and effectiveness of repairs.
\newblock In \emph{CHI '5}, pages 31--40, 2005.
\newblock ISBN 1-58113-998-5.

\bibitem[Borlund(2003)]{Borlund:2003}
P.~Borlund.
\newblock The {IIR} evaluation model: a framework for evaluation of interactive
  information retrieval systems.
\newblock \emph{Information research}, 8\penalty0 (3), 2003.

\bibitem[Dehghani et~al.(2017)Dehghani, Jagfeld, Azarbonyad, Olieman, Kamps,
  and Marx]{Dehghani:2017}
M.~Dehghani, G.~Jagfeld, H.~Azarbonyad, A.~Olieman, J.~Kamps, and M.~Marx.
\newblock Telling how to narrow it down: Browsing path recommendation for
  exploratory search.
\newblock In \emph{Proceedings of the CHIIR '17}, 2017.

\bibitem[Diriye et~al.(2010)Diriye, Wilson, Blandford, and
  Tombros]{Diriye:2010}
A.~Diriye, M.~L. Wilson, A.~Blandford, and A.~Tombros.
\newblock Revisiting exploratory search from the {HCI} perspective.
\newblock In \emph{HCIR 2010}, pages 99--102, 2010.

\bibitem[Hsieh-Yee(2001)]{HsiehYee:2001}
I.~Hsieh-Yee.
\newblock Research on web search behavior.
\newblock \emph{Library \& Information Science Research}, 23\penalty0
  (2):\penalty0 167--185, 2001.

\bibitem[Kelly(2009)]{Kelly:2009}
D.~Kelly.
\newblock Methods for evaluating interactive information retrieval systems with
  users.
\newblock \emph{Found. Trends Inf. Retr.}, 3\penalty0 (1--2):\penalty0 1--224,
  2009.
\newblock ISSN 1554-0669.

\bibitem[Kelly et~al.(2015)Kelly, Arguello, Edwards, and Wu]{Kelly:2015}
D.~Kelly, J.~Arguello, A.~Edwards, and W.-c. Wu.
\newblock Development and evaluation of search tasks for {IIR} experiments
  using a cognitive complexity framework.
\newblock In \emph{ICTIR '15}, pages 101--110, 2015.

\bibitem[Olieman et~al.(2014)Olieman, Azarbonyad, Dehghani, Kamps, and
  Marx]{Olieman:2014}
A.~Olieman, H.~Azarbonyad, M.~Dehghani, J.~Kamps, and M.~Marx.
\newblock Entity linking by focusing on dbpedia candidate entities.
\newblock In \emph{ERD '14}, pages 13--24, 2014.

\bibitem[White and Drucker(2007)]{White:2007}
R.~W. White and S.~M. Drucker.
\newblock Investigating behavioral variability in web search.
\newblock In \emph{WWW '07}, pages 21--30, 2007.

\bibitem[White and Roth(2009)]{White:2009}
R.~W. White and R.~A. Roth.
\newblock Exploratory search: beyond the query-response paradigm.
\newblock \emph{Morgan and Claypool Publishers}, 3, 2009.

\bibitem[Wilson et~al.(2010)Wilson, Kules, Shneiderman, et~al.]{Wilson:2010}
M.~L. Wilson, B.~Kules, B.~Shneiderman, et~al.
\newblock From keyword search to exploration: Designing future search
  interfaces for the web.
\newblock \emph{Foundations and Trends in Web Science}, 2\penalty0
  (1):\penalty0 1--97, 2010.

\end{thebibliography}

\end{document}